\documentclass[aps,twocolumn]{revtex4}
\global\parskip 6pt
\newcommand{\be}{\begin{eqnarray}}
\newcommand{\ee}{\end{eqnarray}}

\newcommand{\lb}{\label}
\def\>{\rangle}
\def\<{\langle}

\def\be{\begin{eqnarray}}
\def\ee{\end{eqnarray}}
\def\lb{\label}
\begin{document}
\title{Horizon State, Hawking Radiation and Boundary Liouville Model}
\author{
Sergey N. Solodukhin
}
\affiliation{{\it School of Engineering and Science, 
International University Bremen,
 P.O. Box 750561, Bremen  28759, Germany}}

\begin{abstract}
\noindent {We demonstrate that the near-horizon physics, the
Hawking radiation and the reflection off the radial potential barrier, 
can be understood entirely within a conformal field theory picture in 
terms  of one- and two-point functions in the boundary Liouville theory. 
An important element in this demonstration is the notion of 
{\it horizon state}, the Hawking radiation being interpreted
as a result of the transition  of  horizon state to the ordinary states 
propagating outside black hole horizon.

}
\end{abstract}
\vskip 2.pc
\maketitle

\noindent {\it Introduction.} 
For almost  three decades, since it was discovered \cite{Hawking:sw}, the 
Hawking radiation  remains one of the mysterious 
and poorly understood phenomena which place
within the intrinsically unitary quantum mechanical picture is still widely debated.
Ultimately it is related to the proper understanding of the
thermodynamical nature of black holes the problem of the entropy
lying in the heart of this challenging issue.
It has been long suspected
\cite{Carlip:1994gy}-\cite{Sachs:2001qb}
that the essential
features of black hole (the entropy, Hawking radiation and
grey-body factors) should have a conformal field theory description.
Examples in string theory \cite{Strominger:1996sh}-\cite{Maldacena:1997ih}
confirm this expectation for certain types of extreme (and
near-extreme) black holes. On the other hand, the general arguments can be given 
\cite{Strominger:1997eq}-\cite{Sachs:2001qb}
for the universal appearance  of the conformal symmetry
in the near-horizon region of a generic stationary black hole.
This symmetry should play an important role in  explaining the black hole 
phenomena.
 
In this note we make a step in this direction and show that
the near-horizon phenomena such as the Hawking radiation and the 
reflection off the potential barrier outside the horizon
have a conformal field theory (CFT) interpretation in terms of one- and two-point
functions in the boundary Liouville field theory (LFT) \cite{Zamolodchikov:1995aa}, 
\cite{Fateev:2000ik}. That the Liouville model may give a
description for the production rate and the S-matrix of strings
in a time-dependent background was suggested in \cite{Gutperle:2003xf}
(see also \cite{Schomerus:2003vv}).  Here we generalize this idea to a
stationary background given by the near-horizon geometry of the Schwarzschild
black hole (a further generalization to arbitrary  non-extreme black
hole is straightforward).  Approaching this, 
we introduce a notion of  {\it horizon state} which is analogous to the
boundary state in the two-dimensional models. Semiclassically it
can be thought as a state living at the horizon and in this respect differs 
from the ordinary propagating states.
The Hawking effect then can be understood as  transition
between a horizon state and the states which  may propagate up to  infinity. 
The notion of  horizon state should be also important for proper understanding
the CFT based entropy calculation \cite{Carlip:1998wz},
\cite{Solodukhin:1998tc} and for solving the unitarity problem.

\noindent {\it The near-horizon limit.} 
The 4-dimensional Schwarzschild black hole is described by
metric
\be
&&ds^2=-f(r)dt^2+f(r)^{-1}dr^2+r^2d\omega^2~,\nonumber \\
&&f(r)=1-a/r
\lb{1.1}
\ee
where $a=2GM$ is the horizon radius. The inverse Hawking temperature 
$T^{-1}_H=2\pi\beta_H$ is expressed in terms of the horizon radius via
relation $\beta_H=2a$. A (bosonic) field propagating
on this background is described by a field equation which after
separating the angle part $Y_{lm}(\theta)$ and the time-dependent part
$e^{-i\omega t}$ at certain frequency $\omega$
reduces to an effectively one-dimensional wave equation
\be
\left[-{d^2\over dz^2}+V[r(z)]-\omega^2 \right] \psi(z)=0~~,
\lb{1.2}
\ee
where a new coordinate $z=\int {dr\over f(r)}=r-a+a\log (r/a-1)$
is introduced which maps the region $a\leq r< \infty$ into the infinite interval
$-\infty < z < +\infty$. In terms of coordinate $z$ the black hole
metric (\ref{1.1}) is conformal to the so-called optical metric
which is relevant for the description of the wave propagation
near horizon.
The effective radial potential in (\ref{1.2}) takes the form
\be
V(r)=f(r)\left({l(l+1)\over r^2}+{(1-j^2)a\over r^3}\right)~~,
\lb{1.4}
\ee
where $j$ is  spin of the bosonic field (possible values of spin are
$j=0,1,2$).
In this note we are  interested in a limit when the size $a$
of the black hole is large so that the black hole 
metric (\ref{1.1})
can be approximated by the Rindler space-time or its modification. 
For the wave equation this boils down to expanding the radial potential
$V(z)$ in powers of $e^{z/a}$ and keeping few first terms in (otherwise
infinite) series.
Restricting to the first two terms in the 
series we get 
\be
V(z)\simeq {1\over a^2}\left(\beta^2 e^{z/a}-\lambda^2\, e^{2z/a}\right)~,
\lb{1.10}
\ee
where $\beta$ and $\lambda$ are certain functions of  $l$ and $j$.
The first term in (\ref{1.10}) is the contribution of the Rindler space
while the second term is a correction due to deviation of the spacetime 
from that of the Rindler. In terms of angular momentum $l$ we have that
${\lambda\over\beta^2}\sim {1\over l}$ for large $l$ and the near-horizon
expansion of the radial potential can be thought (after shifting 
$z\rightarrow z-2a\ln\beta$)  as  expansion in powers of $1/l^2$.

We analyze the pure Rindler case first.
The radial equation (\ref{1.2}) in this case
is particularly simple
\be
\left(-{d^2\over dz^2}+{1\over a^2}\beta^2 e^{z/a}-\omega^2\right)\, \psi_\omega=0
\lb{1.11}
\ee
and the solution is a Bessel function. Choosing the decaying
for positive $z$ function we get
\be
\psi_\omega(z)={2\beta^{-2ia\omega}\over \Gamma (-2ia\omega )}\,
K_{2ia\omega}(2\beta e^{z/2a})~~,
\lb{1.12}
\ee
where the normalization factor is chosen to make the modes (\ref{1.12})
delta-function normalizable.
Close to the horizon  (region of negative infinite $z$)
the exponential potential in (\ref{1.11}) becomes negligible
and the solution is a combination of left- and right-moving plane-waves.
We then find that
\be
&&\psi_\omega(z)\sim \left(e^{i\omega z}+S(\omega)e^{-i\omega z}\right)\, \, , \,
\nonumber \\
&&S(\omega)=\beta^{-4ia\omega}{\Gamma(2ia\omega)\over \Gamma(-2ia\omega)}~,
\lb{1.14}
\ee
where $S(\omega)$ has the meaning of the reflection 
amplitude for a wave scattered off the exponential well.
Note that $|S(\omega)|^2=1$ so that no transmission occurs. 
This shows that the Rindler space
can not be used to compute the grey-body factor due to transmission
through the potential barrier of a Hawking particle emitted from
horizon. For that we need to make a further step.

\noindent {\it One step beyond Rindler.} 
The first correction to the pure Rindler case is given by the second term
in the radial potential (\ref{1.10}).
The radial differential equation 
\be
\left[-{d^2\over dz^2}+{1\over a^2}(\beta^2 e^{z/a}-\lambda^2 e^{2z/a})-\omega^2\right]
\Phi_\omega (z)=0
\lb{1.20}
\ee
in this case has a solution in terms of confluent hypergeometric functions
which at $z\rightarrow +\infty$ behave as in- and out-going waves,
$e^{\pm i\lambda e^{z/a}}$. 
Considering the solution which is only out-going 
for infinite positive $z$
\be
\Phi_\omega (z)=C(\omega)e^{-z/2a}W_{-{i\beta^2\over 2\lambda},\, i\omega a}(-2i\lambda e^{z/a})~~.
\lb{1.23}
\ee
We should note that changing value of $\lambda$ from zero to infinity
the solution (\ref{1.23}) evolves from the Bessel function (\ref{1.12})
at $\lambda=0$ to the Bessel function $H^{(2)}_{i\omega a}(\lambda e^{z/a})$
for large values of $\lambda$. 
We are now interested in the behavior of the solution (\ref{1.23}) 
close to horizon ($z\rightarrow -\infty$). Using the series expansion of the 
function $W_{\kappa,\mu}(x)$ at small values of $x$ we find that
\be
&&\Phi_\omega (z)\simeq \left(A(\omega)
e^{i\omega z}+B(\omega)
e^{-i\omega z}\right)\nonumber \\
&&A(\omega)=(-2i\lambda)^{{1\over 2}+i\omega a}
{\Gamma (-2i\omega a)\over \Gamma({1\over 2}+{i\beta^2\over 2\lambda}-i\omega a)}\nonumber \\
&&B(\omega)=(-2i\lambda)^{{1\over 2}-i\omega a}
{\Gamma (2i\omega a)\over \Gamma({1\over 2}+{i\beta^2\over 2\lambda}+i\omega a)}
\lb{1.24}
\ee
The reflection amplitude given by the ratio 
$S_\lambda (\omega)=B(\omega)/A(\omega)$
takes the form
\be
S_\lambda (\omega)=(-2i\lambda)^{-2i\omega a} {\Gamma(2i\omega a)\over 
\Gamma(-2i\omega a)}{\Gamma({1\over 2}+{i\beta^2\over 2\lambda}-i\omega a)\over
\Gamma({1\over 2}+{i\beta^2\over 2\lambda}+i\omega a)}
\lb{1.25}
\ee
The quasi-normal modes are defined in this case in usual way
and given by condition $A(\omega)=0$. They come from the poles 
of the Gamma function $\Gamma({1\over 2}+{i\beta^2\over 2\lambda}-i\omega a)$
and fall into the set
\be
\omega_n=2\pi T_H \beta^2\lambda^{-1}-4\pi T_H i (n+1/2)~,~~n\in Z~~.
\lb{1.27}
\ee
Notice that $\Re\, \omega_n\sim l$ for large angular momentum $l$.
The imaginary part of $\omega_n$ does not depend on
$\beta$ or $\lambda$ (i.e. $l$ and $j$) and is universal.
For real values of frequency $\omega$ the quantity $|S_\lambda (\omega)|^2$
has a simple form
\be
|S_\lambda 
(\omega)|^2=e^{-2\pi\omega a}{\cosh \pi (\omega a+{\beta^2\over 2\lambda})\over 
\cosh \pi (\omega a-{\beta^2\over 2\lambda})}~~.
\lb{1.28}
\ee
The transmission coefficient $|T|^2=1-|S|^2$ is non-vanishing
in this case so that the flux of the Hawking radiation thermally emitted from 
horizon and escaping through the potential barrier to infinity reads
\be
&&{\cal F}_\lambda (\omega) \sim {|T(\omega)|^2\over e^{\omega \over T_H}-1}
={e^{-{\omega \over 2T_H}} \over e^{\omega\over 2T_H}+e^{\pi\beta^2\over\lambda}}~~.
\lb{1.30}
\ee
This expression should be compared to the ones given in
\cite{Maldacena:1996ix}, \cite{Maldacena:1997ih}.
 
\noindent {\it Horizon state and the Hawking radiation.} Our discussion so 
far was in pretty much standard terms. 
But now  we are going to introduce a new object, {\it horizon state},
which is analogous to the boundary state in two-dimensional
integrable models.
Semiclassically, we define the horizon state $\phi_H$
by  condition
\be
\lim_{z\rightarrow -\infty}
\left(\partial_z +{i\mu_H\over 2a}e^{z/ 2a}\right)
\phi_H =0~~.
\lb{2.1}
\ee
It should be considered as asymptotic 
condition valid for $z$ lying within the interval $-\infty < z < L$
in the close vicinity of the horizon. So that the function satisfying
(\ref{2.1}) takes the form 
\be
\phi_H(z)=e^{-i\mu_H e^{z/  2a}}
\lb{2.2}
\ee
inside the interval and vanishes otherwise.
The function (\ref{2.2}) then indeed describes the mode which
lives only in the small neighborhood of the horizon
and thus differs from the usual {\it propagating} modes
which live everywhere outside horizon and are able to escape to
the asymptotic infinity.
Condition (\ref{2.1}) can be re-formulated in terms of the Rindler coordinate
$\rho$ (defined as $\rho^2=f(z)\simeq e^{z/a}$) as
$
\left( \partial_{\rho}+{i\mu_H\over 2a}\right)_{\rho\rightarrow 0}
\phi_H=0
$
so that $\mu_H$ can be thought as a ``momentum'' conjugate to the 
coordinate $\rho$ as measured at the horizon. 
Instead of using the finite interval
parameterized by $L$ it is convenient to add some imaginary part,
$\mu_H-i\epsilon$, to the momentum and extend the function
(\ref{2.2}) to all values of $z$. It nevertheless
falls off rapidly with $z$ and is sharply peaked at $z=-\infty$
where it approaches a constant.
The latter signals that horizon state is   not normalizable. 
However, it has a non-trivial overlap with the propagating
modes normalizable on the delta-function. In the nearest neighborhood of the
horizon we introduce such a propagating mode as
\be
\lim_{z\rightarrow -\infty} \phi_\omega(z)=e^{i\omega z}~~
\lb{2.5}
\ee
with the usual delta-function normalization.
For positive $\omega$ the mode (\ref{2.5})  describes an out-going wave 
while switching 
the sign of $\omega$ we get an in-going near horizon wave.
The overlap of the mode (\ref{2.5}) with  horizon state (\ref{2.2}) 
is given by the integral
\be
u(\omega)=\int_{-\infty}^{+\infty}dz \,
e^{-i\omega z}e^{-(\epsilon+i\mu)e^{z/2a}}\nonumber \\
=2a{e^{-\pi a\omega}\over (\mu_H-i\epsilon)^{-2ia\omega}}
\Gamma (-2ia\omega)~~.
\lb{2.6}
\ee
So that the horizon state (\ref{2.2}) can be Fourier decomposed in terms of the
modes (\ref{2.5}) as follows
\be
\phi_H(z)={1\over 2\pi}\int_{-\infty}^{+\infty}d\omega \, \phi_\omega(z) \, u(\omega)~~.
\lb{F}
\ee
The function $u(\omega)$ thus completely defines a given {\it horizon state}.
The latter can be considered as a certain superposition of in- and out-going 
modes near horizon.
It is interesting to compute the absolute square of $u(\omega)$.
Assuming that $\mu>0$ and $\omega>0$ we find that
\be
|u(\omega)|^2={1\over T_H\omega}\,{1\over e^{\omega\over T_H}-1}\, ,
\lb{2.7}
\ee
where we take $\epsilon\rightarrow 0$.
It is not difficult to recognize in (\ref{2.7})
the emission probability for the thermal radiation with Hawking temperature 
$T_H=(4\pi a)^{-1}$.
The switching  the sign of $\omega$ in (\ref{2.6}), (\ref{2.7})  would give us 
an absorption probability.
The overlap computed in (\ref{2.6}) can be thought as an amplitude for the
transition of the horizon state to an out-going propagating state
described near horizon by the wave function $e^{i\omega z}$ 
which further scatters off the potential barrier 
and may eventually escape to  the asymptotic infinity.

In the Rindler approximation considered above the propagating mode 
is given by function (\ref{1.12})
and contains both in- and out-going modes near horizon.
Therefore its overlap with the horizon state (\ref{2.2})
contains both the thermal emission  with the Hawking factor as well as
the thermal absorption. This is seen from  the integral
\be
&\int   \psi_\omega(z)\phi_H(z)
=(u_{\tt abs}(\omega) e^{2ia\omega s}+u_{\tt em}(\omega) e^{-2ia\omega s})\,
, & \nonumber \\
&u_{\tt a(e)}(\omega)=2a\beta^{-2ia\omega}e^{\pm\pi a\omega}\Gamma (2ia\omega)\, , 
\lb{2.9}
\ee
where $- (+)$ under the exponent stands for the emission (absorption) 
amplitude and   $s$ is defined as
$
\sinh s=-\mu_H /(2 \beta) .
$
Notice the relation between the emission and absorption amplitudes in (\ref{2.9})
\be
u_{\tt abs}(\omega)=S(\omega)u_{\tt em}(-\omega)~~,
\lb{2.13}
\ee
and the reflection amplitude $S(\omega)$ defined in (\ref{1.14}).

\noindent {\it The near-horizon (boundary) Liouville model.} 
The  just considered  semiclassical picture of the
near-horizon physics, i.e. the propagating modes with reflection amplitude
$S(\omega)$
and the horizon state with the Hawking type decay into a propagating state,
can be consistently embedded into a Conformal Field Theory picture
and thus have  entirely conformal description. 
We now consider an extended object, string, the near-horizon dynamics of which 
is described by the Liouville model
(see also \cite{Frolov:1999my} for LFT in  a related context)
 on a disk with certain conformal invariant boundary interaction. 
The LFT action is 
\be
&&A=\int_{\cal D} \left( {1\over 4\pi \alpha'}(\partial Z)^2 +\mu e^{Z/a}\right)
d^2x \nonumber \\
&&+\int_{\partial {\cal D}} \left({q k\over 2\pi \alpha'}Z +i\mu_H e^{Z/2a}\right) d\xi~~,
\lb{3.1}
\ee
where $k$ is the extrinsic curvature of the boundary $\partial {\cal D}$.
The domain $\cal D$ can also be the upper half-plane.
In the stringy picture the Liouville field $Z$ can be  understood as a 
radial coordinate of the string in the near-horizon optical metric.
In the Hawking type process, for 
a distant observer, strings are emerging from the horizon at certain rate.
To model this we take a time-like boundary staying arbitrary close
to the horizon and consider string configurations which may end on this boundary.
(The latter thus can be treated in a way similar to D-brane.)
This motivates the including in (\ref{3.1}) the contribution of
boundary of the string worldsheet with the boundary terms completely fixed by
the conformal invariance and the correspondence to the field theory limit, as
we show below.
The relevant to our purposes results on quantum Liouville model
can be found in \cite{Fateev:2000ik}. In what follows we re-formulate 
them in slightly different terms more appropriate in view of the action
(\ref{3.1}). This action can be brought to the standard Liouville form 
with exponential bulk potential $e^{2b\phi_L}$ and boundary potential
$e^{b\phi_L}$ in terms of dimensionless field $\phi_L(x)$ (see  \cite{Fateev:2000ik})
by substitution
\be
Z=\sqrt{\alpha'}\phi_L \, , \, b=\sqrt{\alpha'}/(2a)~~.
\lb{3.2}
\ee
The quantity $q$ in (\ref{3.1}) is related to the usual background charge 
$Q=q/\sqrt{\alpha'}=b+{1\over b}$
in the Liouville model. It determines the central charge of the theory $c_L=1+6 Q^2$.
Recall that $a$ is the size of the black hole so that to the leading order in
$\alpha'$ the central charge $c_L$ is proportional to the horizon area, similarly to the
near-horizon CFT considered in \cite{Carlip:1998wz}, \cite{Solodukhin:1998tc}.
The primary operators in LFT are exponential operators
\be
V_\omega (x)=e^{(q/\alpha'+i\omega)Z}
\lb{3.4}
\ee
with conformal weight $\Delta_\omega=Q^2/4+\alpha'\omega^2/4$
(notice that the relation between parameter $P$ used in \cite{Fateev:2000ik}
and our parameter $\omega$ is given by $2P=\sqrt{\alpha'}\omega$).
Before further analyzing the model (\ref{3.1}) let us consider its
semiclassical limit. It is the limit when $\alpha'\rightarrow 0$
($b\rightarrow 0$). In this limit the minisuperspace approximation
in which only dynamics of the zero mode $Z_0$ is taken into account
while the oscillator modes are neglected is expected to work.
In the minisuperspace approximation the primary
state $|v_\omega>$  is represented by the wave function 
$\psi_\omega (Z_0)$ which satisfies
equation
\be
(-{d^{2}\over dZ^{2}_0}+4\pi \mu e^{Z_0/a}-\omega^{2})\psi_\omega(Z_0)=0~~.
\lb{3.5}
\ee
It thus reproduces (after making the identification $4\pi\mu={\beta^{2}\over a^{2}}$)
the field theory wave function (\ref{1.12}) in the Rindler space.
The other observation is that the boundary state wave function
which in the minisuperspace approximation is exponential of the boundary Lagrangian
is in fact  given by (\ref{2.2}) and thus is semiclassically identical to
what we call {\it horizon state}.
Thus in the field theory limit the Liouville model (\ref{3.1}) is consistent
with the near-horizon analysis given in the first part of this note.
Of our further main interest are the bulk one- and two-point functions for the
operators (\ref{3.4}). The bulk one-point function
\be
<V_\omega (x)>={U(\omega,\mu_H)\over |x-\bar{x}|^{2\Delta_\omega}}
\lb{3.5'}
\ee
can be also interpreted as the matrix element between 
a primary physical state $|v_\omega>$ (created by the operator (\ref{3.4})) 
and the boundary state.  
The boundary state thus can be viewed as a collection of the bulk one-point functions.
In our near-horizon picture we identify the boundary state in the model
(\ref{3.1}) with the horizon state $|H>$ semiclassically given by the wave
function (\ref{2.2}).
Thus, we have that $U(\omega,\mu_H)=$ \\ $<H|v_\omega>$ and 
adapting the expression given in  \cite{Fateev:2000ik}
to our case (notice that importantly the boundary coupling is 
imaginary in (\ref{3.1})) we find that
\be
U(\omega)&=&U_{\tt abs} (\omega)e^{2ia\omega s}+U_{\tt em}
(\omega)e^{-2ia\omega s}\, ,
\lb{3.6}\\
U_{\tt a(e)}
(\omega)&=&
i\alpha'\omega (\pi\gamma({4a^2\over \alpha'}))^{-ia\omega}
\Gamma({i\alpha'\omega\over 2a})\Gamma(2ia\omega)e^{\pm \pi a \omega}\nonumber
\ee
where $\gamma(x)=\Gamma(x)/\Gamma(1-x)$ and $s$ is related to the parameters 
of the bulk and boundary coupling
as $\sinh^2 s={\mu^2_H\over \alpha'\mu}\sin({\alpha'\pi\over 4a^2})$.
This expression is a $\alpha'$-modification of our-semiclassical expression
(\ref{2.9}), quantities $U_{\tt abs}(\omega)$ and $U_{\tt em}(\omega)$ have
natural interpretation as exact (closed string) absorption and emission amplitudes.
Calculating the emission probability we get
a $\alpha'$-modification of the Planck spectrum (\ref{2.7})
\be
|U_{\tt em}(\omega)|^2={1\over T_H\omega}{1\over e^{\omega\over T_H}-1}\,\,
{2\omega\over T^*_H}{e^{\omega\over T^*_H}\over e^{2\omega\over T^*_H}-1}\, ,
\lb{3.7}
\ee
where we have introduced ``dual'' temperature $T^*_H=2a/(\pi\alpha')$.
It is a prediction for corrections to Hawking's formula due to
the finite size of string. 
Clearly, when $\alpha'\rightarrow 0$ the standard expression (\ref{2.7}) is
reproduced.
The two-point function, when both points are far away from the boundary, 
\be
<V_\omega(x) V_\omega(x')>={S_L(\omega)\over |x-x'|^{4\Delta_\omega}}
\lb{3.8}
\ee
is related to the Liouville reflection amplitude 
\be
S_{\tt L}(\omega)=-(\pi\gamma({4a^2\over \alpha'}))^{-2ia\omega}
{\Gamma(2ia\omega) \over \Gamma(-2ia\omega)}{  \Gamma({i\alpha'\omega\over 2a})\over 
\Gamma(-{i\alpha'\omega\over 2a})}
\lb{3.9}
\ee
that is a $\alpha'$-deformation of the semiclassical expression
(\ref{1.14}) in the Rindler space. The relation between exact quantum
absorption and emission amplitudes (\ref{3.6}) and the reflection amplitude
(\ref{3.9}) is given by
\be
U_{\tt abs}(\omega)=S_{\tt L}(\omega) U_{\tt em}(-\omega)
\lb{3.10}
\ee
just as in the semiclassical case (\ref{2.13}). This completes our
demonstration of  a conformal field theory description of the
near-horizon physics, the Rindler reflection amplitude and the Hawking radiation.
The deviation of the spacetime from that of Rindler can be taken into account
by perturbing the conformal Liouville theory (\ref{3.1}) by  a
bulk conformal operator $e^{2Z/a}$ with certain (negative) coupling. 
In the resultant theory the transmission to the region of large positive $Z$,
as we have seen,
is not semiclassically forbidden. So that the two-point correlation
function in the perturbed theory should give us  a conformal
description of the grey-body factor and the flux 
(semiclassically given by (\ref{1.30})) 
as well as the quasi-normal modes (\ref{1.27}).

I would like to thank  M. Baumgartl, J. de Boer, J. Fernandez-Barbon,
V. Frolov, M. Gutperle, K. Krasnov,
G. Kunstatter, R. Mann, A.Mikhailov, M. Parikh, J. Polchinski, I. Sachs and V. Schomerus for useful discussions.
This project was started at LMU, Munich and  supported by
the grant DFG-SPP 1096.

\end{document}